# MODELLING COAGULATION SYSTEMS: A STOCHASTIC APPROACH

V.V. Ryazanov

Institute for Nuclear Research, Pr. Nauki, 47, 03068, Kiev, Ukraine

ABSTRACT

A general stochastic approach to the description of coagulating aerosol system is developed. As the object of description one can consider arbitrary mesoscopic values (number of aerosol clusters, their size etc). The birth-and-death formalism for a number of clusters can be regarded as a partial case of the generalized storage model. An application of the storage model to the number of monomers in a cluster is discussed.

## 1. INTRODUCTION.

The stochastic description of the coagulation process was performed in papers like (Scott,1977; Bayewitz et all, 1971; Lushnikov, 1978; Merculovich, Stepanov, 1985, 1986, 1991). So, the birth-and-death model leads to the kinetic equation of coagulation in the form

$$\partial P(t, [X])/\partial t = (1/2) \sum_{i \neq j} W(i,j) [(X_i+1)(X_j+1)P(X_i+1, X_j+1, X_{i+j}-1) - X_i X_j P] + \quad (1)$$

$$+ (1/2) \sum_i W(i,i) [(X_i+2)(X_i+1)P(X_i+2, X_{2i}-1) - X_i(X_i-1)P] \,,$$

where $P(t, X_1, X_2, ..., X_n, ...)$ is the probability to find $X_i$ particles (clusters) having the size $i$ ($i=1,2,...$) in the time $t$; $W(i,j)$ is the coagulation probability per time unit of the particles $i$ and $j$ (containing, in general, the factor $L^{-3}$, where $L$ is the size of a system). The equation for the generating functional

$$F(t, [s]) = \sum_{[X]} \prod_i (s_i)^{X_i} P(t, [X]) \quad (2)$$

can be drawn from (1):

$$\partial F/\partial t = (1/2) \sum_{i,j} W(i,j) (s_{i+j} - s_i s_j) \partial^2 F/\partial s_i \partial s_j \,. \quad (3)$$

The equation for the average number of clusters (from either (1) or (3)):

$$\partial \langle X_k \rangle / \partial t = (1/2) \sum_{ij} W(i,j) D(i,j|k) Q_2(i,j) \,; \quad (4)$$

$$D(i,j|k) = \delta(i+j;k) - \delta(i;k) - \delta(j;k); \quad Q_2(i,j) = \partial^2 F/\partial s_i \partial s_j |_{[s]=1} = \langle X_i (X_j - \delta(i;j)) \rangle \quad (5)$$

($\delta(i;j)$ is the Cronecker symbol) is unclosed since higher momenta $Q_k$ are involved for which successive set of equations can be derived from (1), (3). If one makes an assertion that the random number of clusters of each size has the independent Poisson statistics then

$$Q_2(i,j) = \langle X_i(X_j-\delta(i;j))\rangle = \langle X_i\rangle\langle X_j\rangle , \qquad (6)$$

and one arrives at the Smolukhovsky equation from (4):

$$\partial\langle X_k\rangle/\partial t = (1/2) \sum_{ij} W(i,j) D(i,j|k) \langle X_i\rangle\langle X_j\rangle . \qquad (7)$$

In (van Dongen, 1987) the transition from (4) to (7) was performed basing on the method of van Kampen (van Kampen, 1984). In papers (Merkulovich, Stepanov, 1991, 1992) the spatially inhomogeneous coagulating systems were treated using the discretization operations both in space and time. The stochastic storage model for the random number of monomers in a cluster was introduced in (Ryazanov, 1991; Ryazanov, Shpyrko, 1994). In the present paper the general stochastic approach for describing arbitrary (random) macroscopic values characterizing an aerosol system is presented. The traditional stochastic storage model is generalized and the results are applied to the investigation of coagulating systems.

## 2. GENERAL STOCHASTIC APPROACH TO THE DESCRIPTION OF COAGULATING AEROSOL SYSTEMS.

The mesoscopic stochastic description is generally meant as an intermediate description level between treating microscopic (molecular) quantities (such as the position and pulse of each molecule) and macroscopic (thermodynamic) ones. The mesoscopic level deals with the distribution function (or stochastic process) for the order parameters whose averages are to be treated macroscopically as thermodynamic quantities. For an aerosol system one could point out such values as number of clusters in a unit volume, size of a given cluster treating them as the order parameters. Denote such an order parameter as q(t) without its concretization for a moment (of cource, q(t) can be readily understood as multycomponent vector as well). In the assumption of the Markovian character of a process the distribution function $\omega(q,t)$ satisfies the master equation of the general type (see, for example, Stratonovich, 1992):

$$\partial\omega(q,t)/\partial t = N_{\partial,q} \Phi(-\partial/\partial q, q) \omega(q,t) , \qquad (8)$$

where the "kinetic operator" $\Phi$ is the contracted notation of the expansion of the Chapman equation; symbol $N_{\partial,q}$ orders the operations of differentiation and those of multiplication by functions of q. Thus the fundamental quantity of the mesoscopic approach is the matrix of transition probabilities (or kinetic operator which is nothing more or less than the generating function on these probabilities). The Laplace transform of the function $\omega(q,t)$

$$F(\exp\{-\theta\},t) = \int_0^\infty \exp\{-\theta y\} \omega_t(y) dy \qquad (9)$$

resembles (2) up to the substitution $s=\exp\{-\theta\}$. The kinetic equation in this representation (that is for *F(exp{-θ})*) is (from (8))

$$\partial F(\exp\{-\theta\},t)/\partial t = N_{\theta,\partial/\partial\theta} \Phi(-\theta,-\partial/\partial\theta) F(\exp\{-\theta\},t) . \qquad (10)$$

We offer some common examples illustrating the specification of the transition matrix (kinetic operator). The diffusion process is by definition

$$\Phi(\theta,q) = K_1(q)\theta + K_2(q)\theta^2/2 \ , \qquad (11)$$

where $K_1$, $K_2$ are drift and diffusion coefficients respectively. Substituting (11) into (8) yields the Fokker-Planck equation. This approximation is quite common in many areas of physics and, *faute de mieux*, it was applied to various effects in the aerosol systems (Fuchs, 1964) - such as the spatial diffusion, filtration, coagulation, sedimentation processes. The value $K_1$ from (11) was thus $V_x$ - the projection of the velocity of the aerosol cluster onto axis x; $K_2/2=D=$const, D is the diffusion coefficient. As the random value q one took the cluster coordinate x (which may be called some external coordinate). The description of a single (separate) cluster was thus performed. A single cluster description was also introduced in (Ryazanov, 1991; Ryazanov, Shpyrko, 1994). As the random number q we took the number m of monomers in a cluster, that is the internal coordinate. The stochastic storage model was used in the kinetic equation for this value

$$dm/dt = dA/dt - r[m(t)] \ , \qquad (12)$$

where A(t) is a random input function, r[m] is the release rate. The input A(t) is given by specifying the cumulant function (Prabhu,1980)

$$E(\exp\{-\theta A(t)\}) = \exp\{-t\ \varphi(\theta)\} \ , \qquad (13)$$

where E(...) means averaging. In the absence of infinitely large jumps of the input flux the cumulant $\varphi(\theta)$ is given by

$$\varphi(\theta) = \int_o^\infty (1-\exp\{-\theta y\})\ \lambda b(y)\ dy \ , \qquad (14)$$

here $\lambda<\infty$ means the intensity of the input jumps, b(x) is the distribution function of the jumps. Thus a number of monomers in a single arbitrary choosen cluster is treated as a random storage in a storage system. For the process (11)-(14)

$$\Phi(-\theta,m)=-\varphi(\theta)+\theta r_\chi(m); \quad r_\chi(m)=r(m)-r(0+)\chi_m; \quad \chi_m=1,\ \text{if}\ m=0, \quad \chi_m=0,\ \text{if}\ m>0\ . \qquad (15)$$

From (8), (15) obtain

$$\partial\omega(m,t)/\partial t = \int_o^\infty (\exp\{-y\partial/\partial m\}-1)\ \lambda_m b_m(y)\ \omega(m)dy + \partial(r_\chi(m)\omega(m))/\partial m \ . \qquad (16)$$

Lets take a pure coagulation process. One traces the fate of an arbitrary choosen cluster supposing that it remains the same in all coagulations, even if it coagulates with larger clusters. Thus the cluster can only grow and only input term in (12) is present (that is r(m) =0). Now make a conjecture as to the shape of $\omega(m)$, $\lambda$, b(x). We assume

$$\omega(m,t) = n(m,t)/N; \quad N = \int_o^\infty n(m,t)dm; \quad \lambda b(x) = \beta(m,x)n(x)/2 \ , \qquad (17)$$

where $n(m,t) = <X_m>/L^3$ is the concentration of clusters with m monomers, $\beta(m,x) = W(m,x)L^3$ is the coagulation coefficient, that is the core of the kinetic coagulation equation. The factor 1/2 arises because one accounts one coagulation act twice. Substituting (17) in (16) at r=0 get

$$(\partial n(m)/\partial t)/N - (\partial N/\partial t)(n(m)/N^2) = \qquad (18)$$

$$=(1/2) \int_o^\infty [\beta(m-y,y)n(y)n(m-y)/N - \beta(m,y)n(y)n(m)/N]dy \, .$$

It is worth while mentioning that the choice of $\omega$ and $\lambda b(x)$ in (17) is not quite correct from the point of view of traditional storage model. There is a dependence of $\lambda b(x)$ in (17) on $\omega(x)$ (through $n(x)=\omega(x)N$), on t (through the time dependence in n(x,t), $\beta(m(t),x(t))$) and on m (through $\beta(m,x)$) which is in contradiction to the primary suppositions of (Prabhu, 1980). Nevertheless we start from (16) supposing that its solution satisfies (17). It was this approximation that led to (18) which in its turn yields the Smolukhovsky equation of the free coagulation

$$\partial \omega(m)/\partial t = N[(1/2)\int_o^m \beta(m-y,y)\omega(m-y)\omega(y)dy - \int_o^\infty \beta(m,y)\omega(y)dy\,\omega(m)] - \omega(m)\partial \ln N/\partial t \,, \qquad (19)$$

if

$$-\partial \ln N/\partial t \cong (1/2) \int_o^\infty \int_o^\infty \beta(x,y)\omega(x)n(y)dxdy \cong (1/2) \int_o^\infty \beta(m,y)n(y)dy \, . \qquad (20)$$

First identity in (20) corresponds to the Smolukhovsky's equation and the second one arises from the fact that one chooses a cluster m on random: whatever cluster of the system can figure instead.

Thus the Smolukhovsky equation (19) is obtained from the general master-equation of the storage theory under following assumptions: a) one supposes in (16) r=0 which corresponds to the situation of the constant growth of a given cluster, b) the solution of (16) should satisfy (17) if r=0, the dependence of $\lambda b(x)$ on m brinding obstacles to the use of the storage model because of c) the cluster m is arbitrary and can be replaced by any other cluster of the system which implies (20).

If one considers, like in (1)-(7) a random number of clusters $X_i$ of the size i the corresponding multycomponent kinetic potential takes on the form

$$\Phi(v_1,v_2,\ldots;X_1,X_2,\ldots) = \sum_{ij} [(\exp\{v_{ij}\}-1)/2] W(i,j) X_i (X_j-\delta(i,j)); \quad v_{ij}=v_{i+j}-v_i-v_j \, . \qquad (21)$$

Substituting (21) into (8) leads directly to (1); into (10) with the transition from (9) to (2) yields (3). As random value q one can take either the cluster energy, velocity, charge etc or several such values simultaneously. In (Fuchs,1964) the backward Chapman equation for the model (11) was applied to a number of important problems in the theory of aerosols (such as the time of diffusive sedimentation etc). This equation was used in (Ryazanov, 1989) for investigating the lifetime of aerosols in the storage model. (Under the notion of lifetime we understand the random time moment of the degeneration of a cluster which itself exists only under certain general conditions of thermodynamic character, such as the existence of stationary states. For example, these conditions break for the case of pure coagulation). The degeneration of all clusters means, for example, the resolving of a cloud. Another outcome of the evolution can be the coagulation of all clusters into a big one which means the precipitation. One more possible general result of the evolution is the transition to the domain of states with no stationarity (Ryazanov, 1993a). The physical manifestation of these effects is either the destruction of a system or at least some phase transition (Ryazanov,1991) when the nature of representative value q changes. It was shown that the storage model seems to fit better for elucidating these occurrences than the Smolukhovsky equation (7), birth-and-death model (1), (21) or diffusion approximation (11). The attainment of zero level by a process can be considered in more general frame of the attainment of an arbitrary level or a border. For the Laplace transform of the probability density g(x,t) for the process q(t) starting at t=0 in the point q(t=0)=x to reach for the first time the point zero at the moment t (that is lifetime of a system equals $\Gamma_x$)

$$E(x,s) = E(\exp\{-s\Gamma_x\}) = {}_0\!\int^\infty \exp\{-st\}\, g(x,t)dt \qquad (22)$$

the relation can be derived from the backward Chapman equation

$$N_{x,\partial}\Phi(\partial/\partial x, x)\, E(x,s) = sE(x,s); \qquad E(x=0,s) = E(x,s=0) = 1 \ . \qquad (23)$$

For average and second moment of $\langle\Gamma_x\rangle = -\partial E(x,s)/\partial s\,|_{s=0}$, $\langle\Gamma^2_x\rangle = \partial^2 E(x,s)/\partial s^2\,|_{s=0}$
one has

$$N_{x,\partial}\Phi(\partial/\partial x, x)\langle\Gamma_x\rangle = -1; \quad N_{x,\partial}\Phi(\partial/\partial x, x)\langle\Gamma^2_x\rangle = -2\langle\Gamma_x\rangle \ . \qquad (24)$$

If an inhomogeneous Markovian process is considered (that is $\Phi(\theta,q,t)$ depends explicitly on time) the equation (23) modifies to include explicitly the initial moment $t_o$

$$s\, E_{to}(x,s) = \partial E_{to}(x,s)/\partial t_0 + s\, N_{x,\partial}\Phi(\partial/\partial x, x, t\to t_o-\partial/\partial s)\,(E_{to}(x,s)/s) \ . \qquad (25)$$

## 3. GENERALIZATION OF THE STORAGE MODEL.

The transform of kinetic potential $\Phi(x,q)$ is written, according to (Stratonovich,1992) as

$$R(y,x) = \int\Phi(y,q)\omega_x(q)dq = \lim_{\tau\to 0}(\tau^{-1})\int[\exp\{y(q_2-q_1)\}-1]\times \qquad (26)$$

$$\times\exp\{-xq\}p(q_2|q_1)\,\omega_{st}(q_1)dq_1 dq_2/F(x) = {}_{n=0}\!\Sigma^\infty y^n\kappa_n(x)/n! \ ,$$

where $p(q_2|q_1)$ are transition probabilities for the Markovian process, $\omega_{st}(q)$ is the stationary distribution, $\omega_x(q)=\exp\{-xq\}\omega_{st}(q)/F(x)$; $F(x)=\int\exp\{-xq\}\omega_{st}(q)dq$; $\kappa_n(x)=\int K_n(q)\omega_x(q)dq$; $K_n(q)=(\tau^{-1})\int(q_2-q_1)^n p(q_2|q_1)dq_2$ are kinetic koefficients. The fluctuation-dissipation relations in nonequilibrium stationary case take on the form

$$R(x,x) = 0 \ . \qquad (27)$$

For the diffusion process (11) $\kappa_n=0$ $n\geq 3$ then $R_D(y,x)=y\kappa_1(x)[1-y/x]$ (as seen from (11) and (27)). For the storage scheme another approximation is used: namely, independence of $\kappa_n(x)$ on x, $n\geq 2$ (or independence on q of $K_n$, $n\geq 2$). Then

$$R_S(y,x) = y\kappa_1(x)\,[1-\kappa_1(y)/\kappa_1(x)]. \qquad (28)$$

In (Stratonovich,1992) the diffusion schema was adopted as basic one to which successive amendments were considered. We develop similar extension procedure for the storage scheme assuming following series for $K_n$:

$$K_n(q) = K_{n,0}+\gamma K_{n,1} q+\gamma^2 K_{n,2} q^2/2!+\ldots+\gamma^k K_{n,k} q^k/k!+\ldots, \qquad (29)$$

where $\gamma$ is the formal expansion parameter. The kinetic potential takes on the form

$$\Phi(y,q)={}_{n=1}\!\Sigma^\infty y^n K_n(q)/n!=yK_1(q)+{}_{k=0}\!\Sigma^\infty(\gamma^k q^k/k!)({}_{n=2}\!\Sigma^\infty y^n K_{n,k}/n!)=yK_1(q)-\eta_o(y)- \qquad (30)$$

$-\gamma\eta_1(y)q - ... - \gamma^k\eta_k(y)q^k/k! - ...;$     $-\eta_k(y) = {}_{n=2}^{\infty}\Sigma y^n K_{n,k}/n!;$     $\eta_0(y) = \varphi(-y) + y\rho;$     $\rho = \partial\varphi(\theta)/\partial\theta \mid_{\theta=0},$

and its transform is

$$R(y,x) = y\kappa_1(x) - \eta_0(y) - \gamma\eta_1(y)\langle A(x)\rangle - ... - \gamma^k\eta_k(y)\langle A^k(x)\rangle/k! - ...;     \langle A^k(x)\rangle = \int q^k \omega_x(q) dq . \quad (31)$$

The expression (31) can be regarded as some series on the basis $1, \langle A(x)\rangle, \langle A^2(x)\rangle, ....,$ which naturally arises from the shape of stationary distribution for concrete case, that is represents the "eigen" basis for the problem. The full series (31) is, of cource, equivalent to the full series of Gaussian scheme, but usually we intend to truncate the series at some $\langle A^k(x)\rangle$; this form (contrary to the usual method implying the truncating at the term $x^k$) seems to be more convenient, for example, for investigating chaotic systems because it is quite natural to investtigate their characteristics (in particular, $K_n(q)$) as arising from some averaging procedure over the areas of parameter space which yields either constant or smoothly varying in q coefficient functions. Applying (27) to (31) get (Shpyrko, Ryazanov, 2006)

$$R_S(y,x) = y\kappa_1(x) [1 - \kappa_1(y)/\kappa_1(x)] - {}_{k=1}^{\infty}\Sigma \gamma^k \varphi_k(y)/k! [\langle A^k(x)\rangle - \langle A^k(y)\rangle] \quad (32)$$

with "arbitrary" $\varphi(x)$. he functions $\varphi$ are thus "dissipative undetermined" (Stratonovich, 1992) in the macroscopic approach of (30)-(32) which is based on taking as primitive (initial) quantities the "observables", that is a) stationary distributions and b) equations of motions (stored, as one can easily check, in $\kappa$). Another approach to the model specification consists in specificating rather the process generating these observables than the observables themselves. Thus we arrive at the specification of the process in terms of jump input and release rates. Split $K_1(q)$ in (30) into two parts $K_1(q) = \rho - r(q)$, where $r(q)$ is arbitrary release function and $\rho(q)$ enters into (30) as the term with n=1. One can write the series analogous to (29): $\rho(q) = \rho_o + \gamma \rho_1 q + \gamma^2 \rho_2 q^2/2! + ...$ , where $\rho_l$ are coefficients $K_{n,k}$ from (29), n=1, k=0,1,2,... In this case we arrive at the kinetic potential analogous to the ordinary storage model (15)

$$\Phi_S(y,q) = -yr(q) + {}_{k=0}^{\infty}\Sigma(\gamma^k q^k/k!)({}_{n=1}^{\infty}\Sigma y^n K_{n,k}/n!) . \quad (33)$$

Functions

$$-\varphi_k(y) = {}_{n=1}^{\infty}\Sigma y^n K_{n,k}/n! \quad (34)$$

can be interpreted (like (14)) in terms of input functions. Generalized input intensity is
$$\lambda(q) = \lambda_o + \gamma\lambda_1 q + \gamma^2 \lambda_2 q^2/2 + ...;     \lambda_k = \varphi_k(y = -\infty) ;$$
distribution function $b(\Delta,q)$ is related with function
$$\varphi = {}_k\Sigma \varphi_k \gamma^k q^k/k! = \lambda - \lambda \int \exp\{-y\Delta\} b(\Delta,q) d\Delta .$$

For example, we can set all functions $\varphi_l(y) = \varphi(y)\lambda_l$ equal within a factor (proportional); this is the situation of the birth-and-death processes. This construction is to some extent analogous to that of (Prabhu, 1994) if one takes the modulating process $I(t)$ (Prabhu, 1994) coinciding with the main process $q(t)$.

Substituting (33) into (10) we obtain the relation

$$\partial F(x,t)\partial t = -x r(d/dx) F - \varphi_0(x) F - {}_{k=1}^{\infty}\Sigma \gamma^k \varphi_k(x) (d/dx)^k F/k! . \quad (35)$$

The equation (35) can be solved in successive approximations in $\gamma$: $F=F_o+\gamma F_1+\gamma F_2/2!+...$
Lets give an example. The expression (12) is rewritten as

$$\Phi(v_1,v_2,...;X_1,X_2,...)=-(1/2)_{ij}\Sigma\ \varphi_{2ij}(v_{ij})\ X_i(X_j-\delta(I,j))\ ;$$

$$\varphi_{2ij}(v_{ij})=_o\!\int^\infty (1-\exp\{-v_{ij}u_{ij}\})\ \lambda_{ij}\ b_{ij}(u_{ij})\ du_{ij}\ ; \tag{36}$$

$$u_{ij}v_{ij}=u_{i+j}v_{i+j}-u_iv_i-u_jv_j;\quad \lambda_{ij}b_{ij}(u_{ij})=\delta(u_i-1)\delta(u_j-1)\delta(u_{i+j}-1)W(I,j)\ ,$$

which coincides with (33) if $\gamma=1$, $r=0$, $k=\delta(k;2)$.
Substituting (36) into (10) leads to

$$\partial F(\exp\{-\theta\},t)/\partial t=-(1/2)_{ij}\Sigma\ \varphi_{2ij}(\theta_{ij})\ (\partial^2/\partial\theta_i\partial\theta_j+\delta(I,j)\partial/\partial\theta_i)\ F(\exp\{-\theta\},t)\ , \tag{37}$$

coinciding with (3) if $\theta=-\ln s_i$, $\partial/\partial\theta_I=-s_i\partial/\partial s_i$.
The Laplace transform of (37) yields $F(\theta,s)=_o\!\int^\infty F(\exp\{-\theta\},t)\exp\{-st\}dt$

$$sF(\theta,s)-F(\exp\{-\theta\},t=0)=(1/2)_{ij}\Sigma W(I,j)(\exp\{-\theta_{ij}\}-1)[\partial^2/\partial\theta_i\partial\theta_j+\delta(I,j)\partial/\partial\theta_i]F(\theta,s)\ .$$

In (Lushnikov, 1978) the generating functional is written in the Smolukhovsky approximation when $X_i \to <X_i>$ and $X_j-\delta(I,j) \to X_j$ in (21) and (36) when the solution for the initial conditions $F(\exp\{-\theta\},t=0)=\exp\{-_i\Sigma\theta_iX_{oi}\}$ has a form

$$F_{sm}(\exp\{-\theta\},t)=\exp\{-_i\Sigma\theta_iX_{oi}-_{ij}\Sigma\varphi_{2ij}(\theta_{ij})_o\!\int^t<X_i(\tau)><X_j(\tau)>d\tau\}\ . \tag{38}$$

The kinetic potential corresponding to the Poisson distribution

$$\Phi_{Sm}(v,X)=_{ij}\Sigma[(\exp\{v_{ij}\}-1)/2]W(I,j)<X_i><X_j>$$

corresponds to the storage model with the input intensity proportional to $<X_i><X_j>$ and zero release. The expressions of (Ryazanov, 1991) for the storage model (12-15) correspond to the approximation

$$\Phi_S(v,X)=_{ij}\Sigma[(\exp\{v_{ij}\}-1)/2]W(I,j)X_{0i}X_{oj}\ ,\quad \text{when}\quad <X_k>=X_{0k}+(t/2)_{ij}\Sigma W(I,j)D(I,j|k)X_{0i}X_{0j}$$

(D is given in (5)). The time $t^*$ of degeneration of the value $N(t)=_k\Sigma<X_k>/V$, $t^*=2N_o/cM^2$, $W(I,j)=cij$, $M=M_o=_k\Sigma k<X_k>/V$ coincides with the results of (Voloschuk, 1984) from the Smolukhovsky equation. Thus, the model suggested first in (Ryazanov, 1991) can be generalized at several levels.

The expressions (36), (37) are readily to yield the equations with higher than (38) precision (involving higher momenta). For example, one gets the refined version of Smolukhovsky equation:

$$\Phi^{(1)}{}_{Sm}(v,X)=_{ij}\Sigma\ (\exp\{v_{ij}\}-1)W(I,j)X_i<X_j>/2$$

$$\partial F^{(1)}{}_{Sm}/\partial t=-(1/2)_{ij}\Sigma\ W(I,j)(\exp\{-\theta_{ij}\}-1)<X_j>\partial F^{(1)}{}_{Sm}(\exp\{-\theta\},t)/\partial\theta_I\ .$$

Rewriting (21) and (36) in the form

$$\Phi(v,X) = {}_{ij}\Sigma\ (\exp\{v_{ij}\}-1)\ [<X_i><X_j>+\Delta_i(<X_j>-\delta(I,j))-$$

$$-\delta(I,j)<X_i>+\Delta_j<X_i>+\Delta_I\Delta_j];\qquad \Delta_I = X_i - <X_i>$$

we get

$$-\partial F/\partial\theta_I\mid_{\theta=0}\ -<X_i> = <X_i> - <X_i> = 0$$

$$\partial<X_k>/\partial t=(1/2)\ {}_{ij}\Sigma W(I,j)D(I,j|k)[<X_i><X_j> + \Delta_i\Delta_j - \delta(I,j)<X_i>]\ ,$$

two last terms representing the amendments to the Smolukhovsky equation.
 From (37) it is possible to derive the "one-particle" Laplace transform of

$$P_k(X_k) = \int...\int \omega(X_1,...,X_{k-1},X_k,X_{k+1},...)dX_1...dX_{k-1}dX_{k+1}...\ ;$$

$$f_k(\exp\{-\theta_k\})= \int...\int \exp\{-\theta_k X_k\}\ \omega(X_1,X_2,...)dX_1 dX_2... = \int \exp\{-\theta_k X_k\}P_k(X_k)dX_k.$$

The equation for f is

$$\partial f_k(\exp\{-\theta_k\})/\partial t = {}_{ij}\Sigma\ (1/2)W(I,j)(\exp\{-\theta_k\}-1)\times$$

$$\times D(I,j|k)\ [<X_iX_j\exp\{-\theta_k X_k\}-\delta(I,j)X_i\exp\{-\theta_k X_k\}>]\ . \tag{39}$$

With an assertion $<X_iX_k\exp\{-\theta_k X_k\}> \cong -<X_i>\ \partial f_k/\partial\theta_k$ ;
$<X_iX_{k-i}\exp\{-\theta_k X_k\}> \cong <X_iX_{k-i}>\ f_k$ (39) yields

$$(\exp\{-\theta_k\}-1)[\partial^2 f/\partial\theta^2 - a\partial f/\partial\theta - fb]\ = -(W(k,k))^{-1}\partial f/\partial t;\quad a=(W(k,k))^{-1}\times$$

$$\times\ [{}_{i=1}^{\infty}{}_{i\neq k}\Sigma W(k,i)<X_i>-W(k,k)];\qquad bW(k,k)={}_{1\leq i\leq k-1}\Sigma(1/2)W(I,k-i)<X_iX_{k-i}-\delta(I,k-i)X_i>$$

with following stationary solution

$$f_{st}(\theta)=[(c-a/2-<X^s_k>)\exp\{(a/2+c)\theta\}+(c+a/2+<X^s_k>)\exp\{(a/2-c)\theta\}]/2c;$$

$$c=[(a/2)^2+b]^{1/2};\qquad <(X^s_k)^2> = b - a<X^s_k>\ .$$

For Poisson distribution
$$<X^2_k> \cong <X_k>^2 + <X_k>\ ,$$

and $<X_k>=[(a+1)/2]^2\pm([(a+1)/2]^2+b)^{1/2}$ .
 Setting
$<X^s_i>=\delta(I,M)$, get $a=w(k,M)/w(k,k)$, $b=0$; $<X^s_k>=w(k,M)/w(k,k)=1$ by $k=M$ .

 The approach of this chapter allows the representation of the birth-and-death and some other kinds of processes in the frame of the generalized storage models.

## 4. APPLICATION OF THE STORAGE MODEL TO THE NUMBER OF MONOMERS IN A CLUSTER.

Consider the random number of monomers m in a single cluster (12)-(20). Combining (15)-(17) with (10) leads to

$$F(\exp\{-\theta\},t)=E(\exp\{-\theta m\})=\int_0^\infty \exp\{-\theta u\}\omega(u,t)du \; ;$$

$$\partial F(\exp\{-\theta\},t)/\partial t=\int[(\exp\{-u\theta\}-1)/2]\beta(u,-\partial/\partial\theta)\, n(u)du F(\exp\{-\theta\},t) \, . \qquad (40)$$

For $\beta(m,x) \approx m^a$ with $1<a<1$ (40) corresponds to fractional terms in the series (29) and fractional derivatives in (40) which is the manifestation of the fractal character of cluster formation (Nigmatullin, 1992). Thus the relations (15-20) used in (Ryazanov, 1991) correspond to the extension of the storage model with generalized Tailor series of fractional derivatives. The explicit form of $\Phi$ in (40) allows us to solve this equation specifying the kernel $\beta(m,x)$. So, for $\beta(m,x)=a=$const taking into account $n(u,t)=N\,\omega(u,t)$ (17) one arrives at (set a=1 without loss of generality)

$$\partial F_a(\theta)/\partial t = a\, (F_a(\theta)-1)\, F_a(\theta)/2 \; ;$$

$$F_a(\theta,t)=y(t)F_a(\theta,t=0)/\{1-F_a(\theta,t=0)[(N(t=0)/2)\int_0^t y(\tau)d\tau+y(t)-1]\};$$

$$y(t)=\exp\{-\int_0^t N(\tau)d\tau\}/2\} \, .$$

Stationary function F=1 corresponds to the delta-shaped peak of the probability density (for example, $\delta(m-M)$). For (17) $\omega(x,t)=n(x,t)/N$, $F(\exp\{-\theta\},t)=n(\theta)/N$, $n(\theta)=\int_0^\infty \exp\{-\theta x\}n(x,t)dx$

These results are in correspondence with the results of (Voloschuk, 1984) for $\beta(m,x)=a=$const. For the kernel $\beta(m,x)=c(m+x)$ (40) get

$$\partial F(\theta)/\partial t = -cN\, [\partial F(\theta)/\partial\theta\, (2F(\theta)-1) + MF(\theta)/N]/2 \, .$$

For $\beta(m,x)=cmx$, $\quad \partial F(\theta)/\partial t=(c/2)N\, [\partial F/\partial\theta + M/N]\, \partial F/\partial\theta$

These equations lead to results which are close to those obtained from the Smolukhovsky equation (Voloschuk, 1984) but do not coincide with them completely. The question remains open about the approximation fitting better to the description of aerosols.

The specification of $\lambda_m b_m(x)$ should lean upon the physical assumptions as to the physics of coagulation. The relation (17) corresponds to the assumption of instantaneous stirring of the system between successive coagulation acts resulting in quasihomogeneous state of the system and thus in the statistical identity of all clusters which (equation (20)) ensures the coincidence of the kinetic equation (16) with the Smolukhovsky equation (19) (if r=0 and (17) holds). One can detalize the model adding some function $b_1(x)$ to (17):

$$\lambda b(x) = \beta(m,x)\, n(x)/2 + \lambda_1 b_1(x) \, . \qquad (41)$$

Then the kinetic equation (16) with the condition (41) and r=0 coincides with (19) (ln N being equal to the first of expressions (20)) under the condition

$$N\int\omega(y) [\int\beta(x,y)\omega(x)dx - \beta(m,y)]dy = 2\lambda_1 \int [(\omega(m-y)/\omega(m))-1] b_1(y)dy , \quad (42)$$

which can be considered as the equation for $\lambda_1 b_1(x)$ given the functions $\beta$ and $\omega$. As $\omega = n/N$ is the required function one arrives at the system of equations (16) with (41)-(42) for two functions $\omega$ and $\lambda_1 b_1$.

When choosing (41) one should add the term $-\varphi_1(\theta)F$; $\varphi_1(\theta) = \int_o^\infty (1-\exp\{-\theta u\})\lambda_1 b_1(u)du$ to the right hand side of (40) that is

$$\partial F(\exp\{-\theta\},t)/\partial t = \int(\exp\{-u\theta\}-1) \beta(u, -\partial/\partial\theta) n(u)du \, F/2 - \varphi_1(\theta) F . \quad (43)$$

So, using (42)-(43) for $\beta=m+x$ (and for other $\beta(x,y)$ as well) we arrive at the expression for $n(\theta)=FN$ coinciding with similar expressions derived from Smolukhovsky equation in (Voloschuk, 1984).

The Smolukhovsky equation is in no manner a standard and we wish to derive an equation yielding more refined description of the aerosol system. So an important task would be the specification of the function $\lambda_1 b_1(x)$ from (41) describing the deviation from the homogeneity of the system after the coagulation event. Lets set $b_1(x)=P_o(t)\delta(x)$ (that is the input of zero clusters) where $P_o(t)$ is the degeneration probability equal to the fraction of time the cluster stays degenerated. From (41) get

$$\lambda b(x) = \lambda_1 P_o(t) \delta(x) + \beta(m,x) n(x)/2 . \quad (44)$$

We can readily set $\lambda_1=\lambda$. The function $b(x)$ is thus proportional to the distribution probability in the storage model which is $\omega(x)=P_o\delta(x)+g(x)$ with $g(x)$ being a continuous function. This is the assumption of the fact that the input to a given cluster consists of other clusters (self-matching storage model). For pure coagulation the stationary probability of degeneration is $P_o=\lim_{t\to\infty}P_o(t)=0$. If $b_1(x)=P_o(t)\delta(x)$, $\varphi_1=0$ (43) takes the form (40). The equation (18) does not change either. Thus we recover the earlier results. But the refinement of the rather crude approximations used above are possible (basing, for example, on the reneval theory) which can contribute to changing (18) and (40). The value $\lambda$ describes the intensity of the cluster motion and thus the frequency of coagulation events. For the monodisperse aerosol with the radius R the expression is written from (Fuchs, 1964) $\lambda=8\pi RDn$, where D is the diffusion coefficient, n is the cluster concentration. Integrating (44) get

$$\lambda_1 P_o=8\pi RDn-\int\beta(m,x)n(x)dx/2. \quad \text{By } \lambda_1=\lambda, \, n(x)=n\delta(x-R), \, \beta(m,x)=8\pi RD, \, P_o=1/2 .$$

Different characteristics of clusters are connected one to another being in fact different aspects of the same problem. For example, the concentration of clusters $n(m,t)$ from (17-20) is expressed through the number of clusters from (1),(21): $n(m,t)=<X_m>/L^3$. This fact allows to proceed from the models like (1), (17) to (15)-(17) (and vice versa) making use of the peculiarities of the behaviour of aerosols found by means of some specific class of models. For example, the external field yields the additional terms in the right hand of equations of (4), (7). When using the equations like (16) these terms are interpreted as some effective release of monomers (plus complication of the problem due to the spatial inhomogeneity). To make a connection between different characteristics of an aerosol system one can use the method of collective variables (van Kampen, 1984), the method of complex generating functional (Ryazanov, 1993b) and identities of the probability theory (Wilde identity etc). Being specified, the kinetic potential allows investigating various temporal characteristics of the aerosol behaviour, for example, by means of expressions (22)-(25).

5. CONCLUSION.

One of the advantages of the stochastic storage model is the clear criteria stated for the existence of stationary states in a system (Ryazanov,1991). Besides the external fields, the stationarity conditions are determined by such factors as the presence of sources and sinks of clusters, evaporation, condensation, splitting etc. For the stationary state of storage model the relations describing its behaviour are known in details. Specific for the aerosol system is the stationary state resulting from the pure coagulation. In this case the release of monomers is zero, and the system is far from being stationary. The input function also vanishes when all clusters form a single bulk cluster; the stationarity conditions remain undefined.

The equations for the distribution function (16) and its Laplace transform (40), (43) encounter serious difficulties coming from the fractional derivatives; these difficulties however reflect the complex fractal (self-similar) behaviour of aerosol clusters. The fractional part of the derivative shows a fraction of systems states conserving during all the time of evolution. In (Nigmatullin, 1992) a broad class of systems was pointed out where one could expect the apparition of the fractional derivatives, coagulating systems belonging to this class as well. Such systems were referred to in (Nigmatullin, 1992) as "systems with residual memory" occupying an intermediate position between full memory systems, on one hand, and completely stochastic (Markovian) ones. Thus the application of purely Markovian processes (like storage or birth-and-death) to the description of coagulation is with necessity an approximation (generally speaking, the properties of a system to be either Markovian or non-Markovian were related in (Nigmatullin, 1992) to the presence of fractional derivatives in time; but the spatial fractional derivative is readily to appear in (40) likewise the first term of the right hand side of equation (16) if $b(x)=cx^{v-1}$, $0<v<1$, $1\leq m\leq M$; in the latter case however the spatial derivative on m can be transferred to the temporal variable and acts on the expression

$$[\partial n(m)/\partial t - (n(m)\partial N/\partial t)/N^2 + c \int_1^M y^{v-1} dy n(m)/2N]) .$$

A generic (typical one) distribution for the storage model is the gamma-distribution (like the Gaussian distribution occupying the same place in the diffusion models). The gamma distribution is known from experiments to fit well the real cluster-size distributions. The storage model is thus open to further detalization of cluster behaviour. In general the stochastic modelling opens vast perspectives for predicting the behaviour of real physical systems. An essential part of the research work in this direction must be the investigation of relations between mathematical abstractions coined by these methods and physical phenomena in a system as it is well known that the real processes in aerosols (as well as in any other natural system) are much more complicated than the mathematical models. Thats why the straightforward use of, e.g. storage models without stepwise and throught analysis of the approximations made is not warranted from faults.